# Cubic Spline Interpolation Segmenting over Conventional Segmentation Procedures: Application and Advantages

Chetan Sai Tutika, Charan Vallapaneni, Karthik R, Bharath KP,
N Ruban Rajesh Kumar Muthu Senior *Member, IEEE*
School of Electronics Engineering
VIT University, India chetansai.tutika@gmail.com

Email: chetansai.tutika@gmail.com,charanvallapaneni@gmail.com
tkgravikarthik@gmail.com,bharathkp25@gmail.com,nruban@vit.ac.in,mrajeshkumar@vit.ac.in

*Abstract*— **To design a novel method for segmenting image using Cubic Spline Interpolation and compare it with different techniques to determine which gives an efficient data to segment an image. This paper compares polynomial least square interpolation and the conventional Otsu thresholding with spline interpolation technique for image segmentation. The threshold value is determined using the above-mentioned techniques which is then used to segment an image into the binary image. The results of the proposed technique are also compared with the conventional algorithms after applying image equalizations. The better technique is determined based on the deviation and mean square error when compared with an accurately segmented image. The image with least amount of deviation and mean square error is declared as the better technique.**

## I. INTRODUCTION

Segmentation of a signal is very important in applications of digital signal processing. In this paper, we use interpolation methods to automatically segment the image [1]. Two different interpolation techniques are used, namely polynomial interpolation and cubic spline interpolation. Using these techniques, we map the minima points from the histogram of the image which we then use to find the threshold and segment the image accordingly.

In polynomial interpolation, the image is broken into several polynomial segments, which are non-overlapping and independent of each other. The methods which are presented can be generalized to higher dimensions, and piecewise-polynomial nature of segments is still valid. Least-squares method is used while interpolating polynomial curve fitting [2][3]. The method of least squares is an approach in regression analysis to approximate the solutions of the systems which have more unknowns than the sets of equations [4].

Although the interpolation theory and calculations are simple, the interpolation effect is not good. Due to this, the spline interpolation algorithm can keep the whole image clarity and smoothness, the cubic spline function has been used widely in many interpolation algorithm [5]. However, the spline interpolation algorithm can create a different degree of edge serrated aging and fuzzification in interpolation process, seriously affecting the image quality. Therefore, it is important the image is post-processed which removes the distortion in the image. Spline is one of the most used interpolation methods for data mapping. The curve fitting function of the extracted points can be acquired using the cubic spline interpolation [6].

## II. SEGMENTATION AND CURVE FITTING

### A. Binary Image segmentation

The term Image segmentation refers to partition into a set of regions that cover it. Meaningful segmentation is the first step from low-level image processing which transforms a grey-scale or color image into one or more other images to high-level image description in terms of features, objects, and scenes. Image processing is dependent on the efficiency of segmentation, but a capable and reliable segmentation of an image is a problem.

Binary Image segmentation is an essential in image processing to split the image into its respective binary form. Segmenting an image to its binary form displays the prominent features of the image. Segmenting is usually carried out with the help of image histogram. The common image property used to threshold an image is pixel grey level: $k(x,y) = 0$ if $p(x,y) < Th$ and $k(x,y) = 1$ if $p(x,y) \geq Th$, where $Th$ is the threshold. Using two thresholds, $Th_2 < Th_1$, a range of grey levels related to region 1 can be defined: $k(x,y) = 0$ if $p(x,y) < Th_1$ OR $p(x,y) > Th_2$ and $k(x,y) = 1$ if $Th_1 \leq p(x,y) \leq Th_2$.

The primary problem with image segmentation is finding an adequate threshold or number of thresholds to separate the desired features from the background. In most practical scenarios, simple thresholding falls short to segment the features of interest.

In this paper, three thresholding methods are detailed which segment the image and results are compared to determine the better technique.

### B. Polynomial curve fitting

Polynomial interpolation is a technique to estimate values between known data points. When the given data contains a gap at few specific points an approximation can be made by using interpolation. The polynomial points are calculated in a least square sense. A process for finding the most accurate-fitting curve to a given set of points is realized by reducing the sum of the squares of the offsets of the points from the curve. The least sum of the squares of the data is considered instead of the offset absolute values which allows the residual as a continuous differential quantity[7].

*C. Cubic Spline interpolation*

Spline interpolation is a type of interpolation in which the interpolant is a form of piecewise polynomial called a spline. Spline interpolation is often used over polynomial interpolation since the interpolation error can be minimized even when using low degree polynomials for the spline[8][9].

Cubic splines are more suitable for curve fitting due to the ease of data-interpolation, differentiation and integration, as they give a smooth response[10][11].

*D. Otsu Thresholding*

Otsu thresholding is a popular algorithm used to segment gray scale image into its respective binary equivalent. The algorithm assumes that the image has two classes of pixels foreground and background pixels, it then finds the threshold separating the two classes so that their combined spread is minimal, and their inter-class variance is maximal.[12]

*E. Curve Fitting*

Curve fitting is a method of constructing a function that has the accurate approximation for a series of data. It may involve either interpolation, where a perfect fit to the data is needed, or smoothing in which a function is constructed that accurately fits the data points. Fitted curves can be used to help with data visualization, to find values of a function where no data is available, and to summarize the co-relation among two or more variables.

III. METHODOLOGY

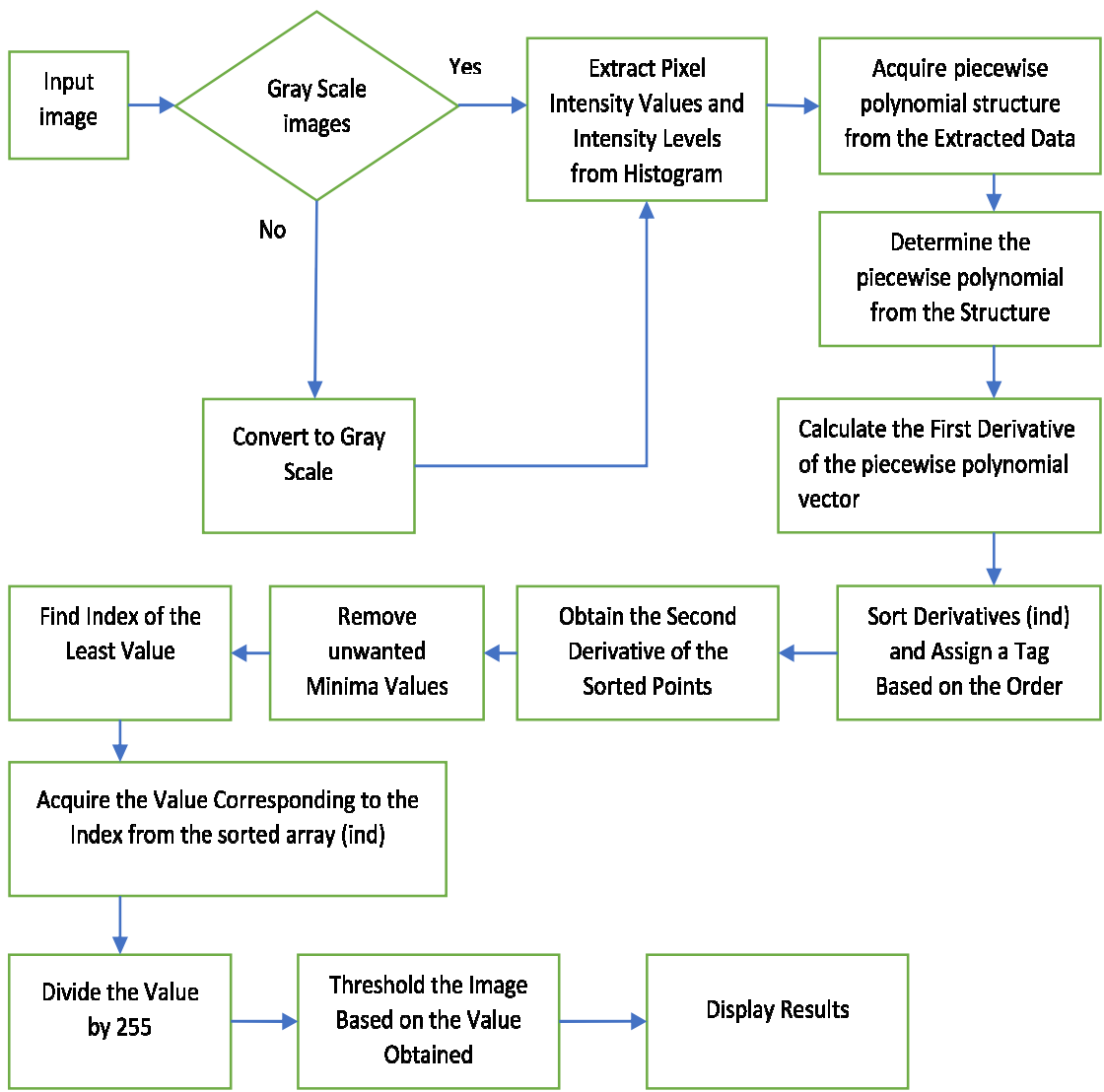

## A. Histogram Acquisition

The image acquired is converted to a gray scale image if required. The histogram of the intensity image is calculated and then displayed. The histogram is split into two variables which hold the values of the intensity (counts) and the bin values which specify the gray levels of the image. The bin locations for gray scale image vary from 0 to 255. Table 1 shows the input image and their respective histograms

Table 1

| Image Name | Input Image(B) | Histogram of the Image(C) |
|---|---|---|
| Rice.png | 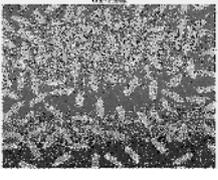 | 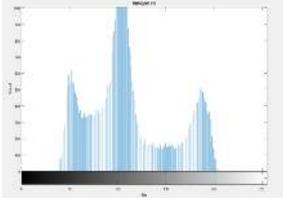 |
| Cameraman.tif | 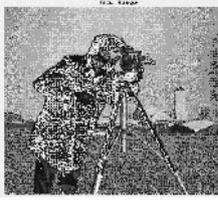 | 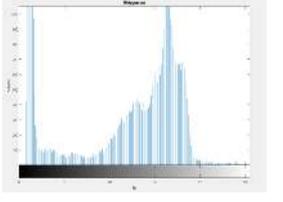 |
| Cell.tif | 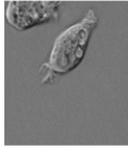 | 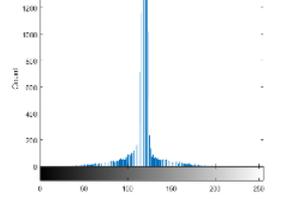 |
| Mri.tif | 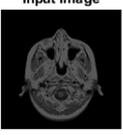 | 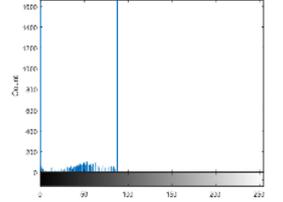 |

Table 1 Figure C Histogram: x-axis =intensity levels(bins); y-axis= pixel intensity values(counts)

## B. Cubic Spline Interpolation

From the Figure C Histograms, the vector of interpolated values and piecewise polynomial structure for the intensity and the bin values are derived. The piecewise polynomial at the bin points is evaluated and mapped against the bin values.

Figure 2 Data Mapped Points on Histogram

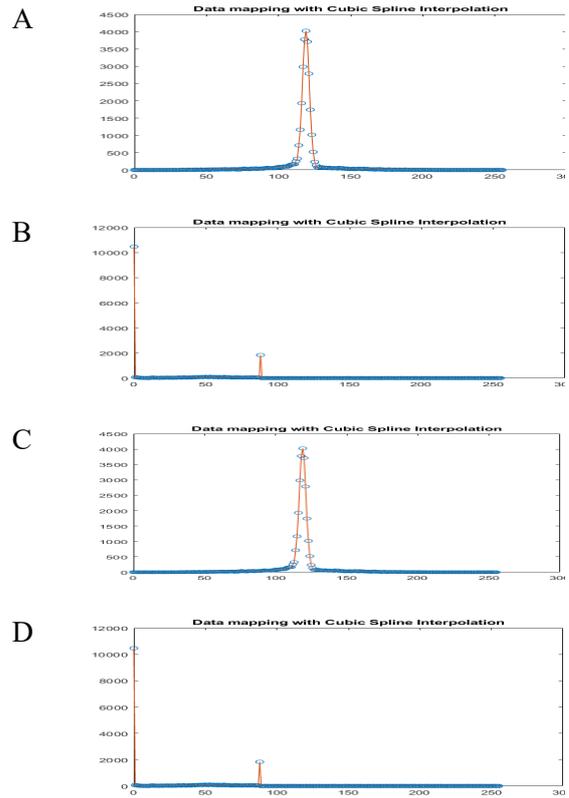

A

B

C

D

Figure 1A cameraman.tif, Figure 1B rice.png Figure 1C cell.tif, Figure 1D mri.tif

## C. Threshold Determination

The curve obtained in Figure 1 Data mapped Points on Histogram and Polynomial curve fitting methods are differentiated to find the zero crossing points of the curve, the absolute values of the difference is taken and arranged in ascending order. To find the minima points of the Figure 1 Data mapped Points, second derivative is applied on Figure 1 Data mapped Points and the minimum points are extracted. In case of Cubic Spline Interpolation multiple minima are obtained at unwanted locations, which are removed by inputting a manual threshold value and eliminating those minima.

The index of the minima is mapped with the index of the sorted first derivative and the respective mapped index value is derived. The value obtained is divided by 255 to get the threshold value.

## D. Post Processing

In case of Cubic Spline Interpolation slight distortions is prevalent in the image. This distortion can be handled by filtering out smaller objects from the segmented images using motion Blur.

## E. Deviaton and Error Calculation

Devaition and Mean square error(MSE) values of the image are used to validate the results and to compare the proposed method with other coventional techniques. Meansquare error in two images is defined as the cumulative square error between the perfectly segmented image and the image segmented using the specified methods.

Deviation=(|PS-IT|)/PS
MSE= $(1/W*H)*\sum\sum(PS-IT)^2$
PS=Accurately Segmented Image, IT=Segmented Image using Stated methods, W,H=Widht and Height of image

## IV. RESULTS

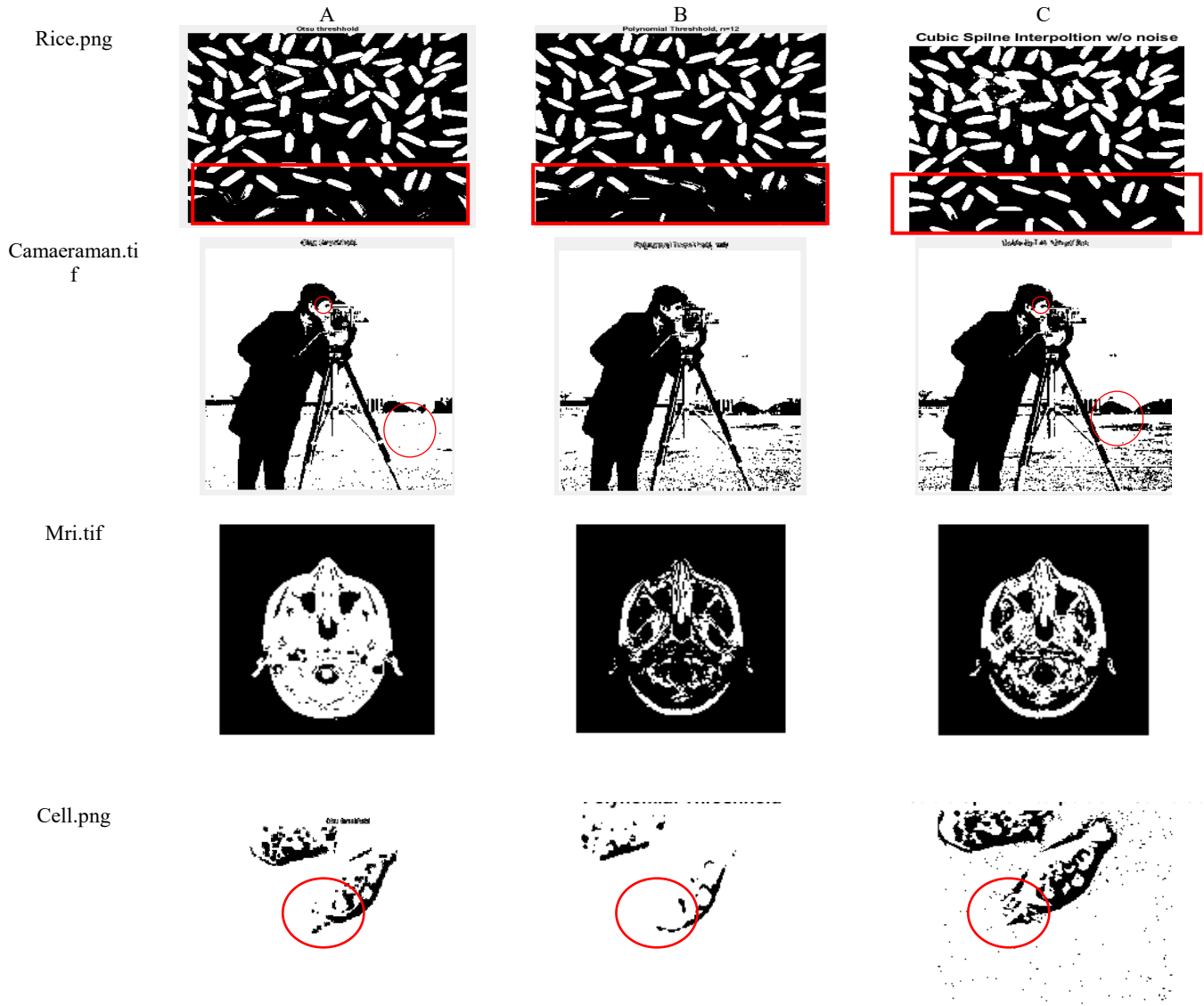

Figure 2A Results by Otsu, Figure 2B Results by Polynomial Curve Fitting, Figure 2C Results by Proposed Method.

From Figure 2, Spline interpolation technique produces better clarity in the segmented image. More features of the image are seen as opposed to the polynomial curve fitting technique and Otsu. While distortions are
prevalent in case of cubic spline interpolation, these can be removed using motion blur operators and other image processing techniques.

Figure 3 Proposed method Comparison with Otsu after Image Histogram and Intensity Adjustments

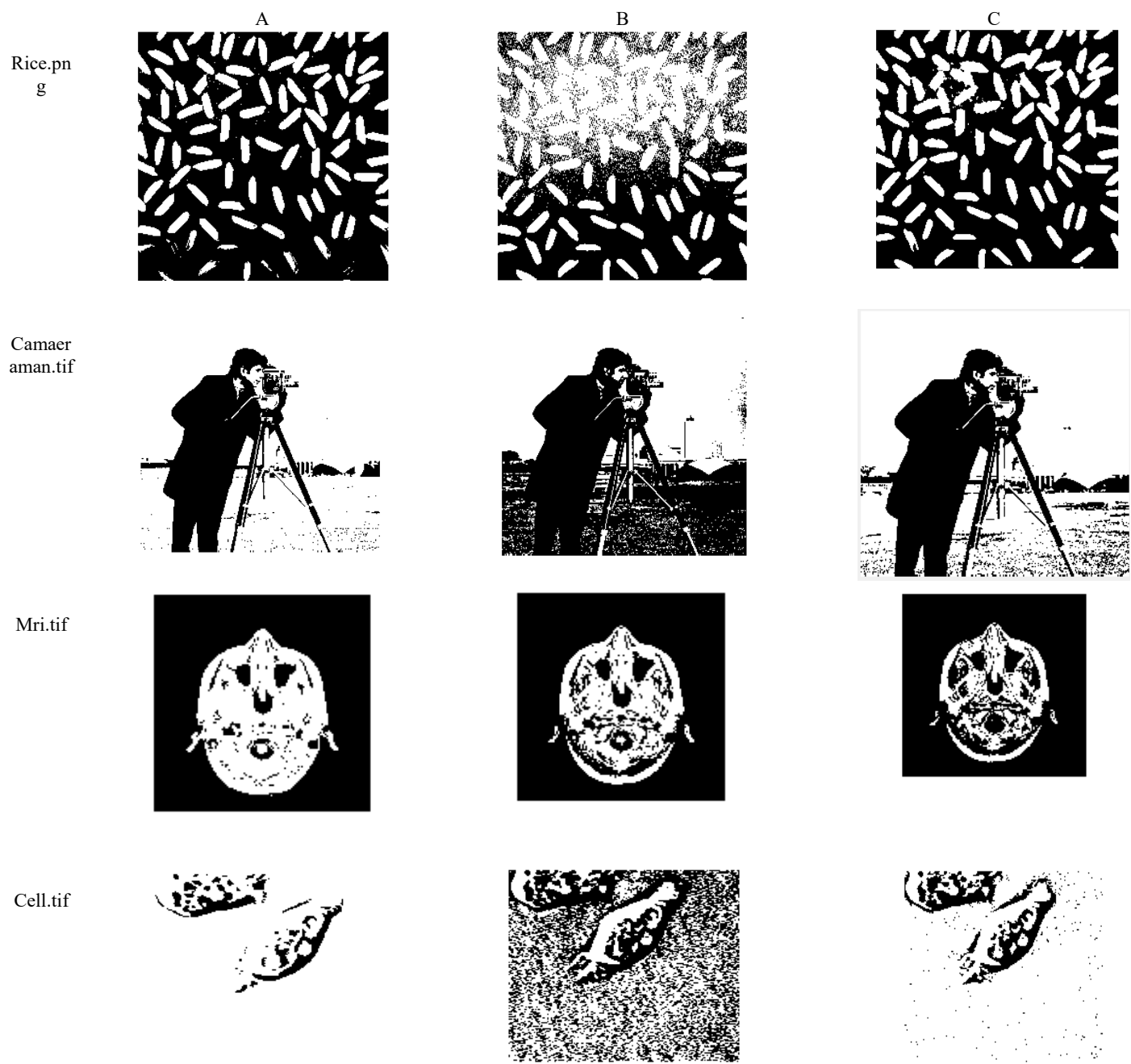

Figure 3A results by Otsu after Intensity Adjustment, Figure 3B results by Otsu after Histogram Equalization, Figure 3C results by Proposed Method

Figure 3 compares the performances of the proposed method with Otsu algorithm after performing histogram equalization and Intensity adjustment. It can be observed from Figure 3 Proposed method Comparison with Otsu after Image Histogram and Intensity Adjustments, that the performance of Otsu is comparable to Cubic Spline Interpolation after performing additional operations. It can also be observed that the proposed method produces better results without any processing required.

Table 2

| Input | Existing Experimental Result[13][14] | Otsu | Polynomial Curve Fitting | Proposed Method |
|---|---|---|---|---|
| Rice.png | 93 | 151 | 172 | 91 |
| Cameraman.tif | 125 | 27 | 49 | 54 |
| Mri.tif | 13 | 2 | 5 | 10 |
| Cell.tif | 29 | 7 | 4 | 24 |

Table 2 depicts the number of contours found in the conventional, proposed methods and compares it to the number of contours obtained with the already accurate segmented image. The Existing Experimental results are obtained after applying image enhancements[13][14] and gives accurate segments.

Table 3

| | Existing Experimental Result[13][14] | | Polynomial Curve Fitting | | Otsu | | Proposed Algorithm | |
|---|---|---|---|---|---|---|---|---|
| Input | Deviation | Mean Square error | Deviation | Mean Square error | Deviation | Mean Square error | Deviation | Mean Square error |
| Rice.png | 0 | 0 | 0.84 | 1.3794e8 | 0.62 | 1.3474e8 | 0.021 | 1.2936e8 |
| Cameraman.tif | 0 | 0 | 0.608 | 4.034e8 | 0.784 | 5.879e8 | 0.568 | 3.937e8 |
| Mri.tif | 0 | 0 | 0.6 | 2.426e8 | 0.84 | 2.797e8 | 0.23 | 1.8481e8 |
| Cell.tif | 0 | 0 | 0.86 | 3.849e9 | 0.75 | 3.66e9 | 0.111 | 3.532e9 |

Table 3 Compares the Deviation and the mean square error obtained when the techniques used in the paper are compared to an accurately segmented image. It can be observed that the deviation and mean square error produced by the proposed method is the least and thus highly accurate.

The piecewise polynomial structure obtained using Cubic spline data interpolation gives the exact approximation of the histogram obtained from the image. While the Polynomial curve fitting returns the best curve fit for the histogram at a least square sense. Polynomial curve fitting may not give the exact approximation for certain cases and might impede the segmentation process. The optimal thresholding value varies in the Polynomial curve fitting technique with respect to the degree of polynomial and in the Cubic spline interpolation with respect to minimal threshold. Even with varying threshold values cubic spline interpolation showed better results in segmenting the Image.

## V. CONCLUSION

In conclusion, Cubic Spline Data interpolation showed better results than Otsu thresholding method and Polynomial curve fitting technique. While noise is added to the image during Cubic Spline Data interpolation, it can be eliminated using motion blur operation and other techniques. By performing Histogram equalization or intensity adjustments on the image, one can get results with Otsu that are comparable to the proposed method. But by using the proposed method one can skip the pre-processing steps to acquire better segmentation results.

The results obtained by the Cubic Spline Interpolation are not entirely noise free. The algorithm can be further modified to get distortion less results without the need of image post processing after segmentation. Developing a more efficient formula to remove the unwanted minima in the image while using Cubic splice interpolation method.